\documentclass[12pt,preprint]{aastex}
\def\degpoint{\ifmmode ^{\rm{o}}\!. \else $^{\rm{o}}\!.$\fi}

\newcommand{\ms}{\mbox{m\,s$^{-1}$}}
\newcommand{\kms}{\mbox{km \ s$^{-1}$}}
\newcommand{\Msun}{\mbox{M$_{\odot}$}}
\newcommand{\Rsun}{\mbox{R$_{\odot}$}}
\newcommand{\Mjup}{\mbox{M$_{\rm Jup}$}}

\newcommand{\Lsun}{\mbox{L$_{\odot}$}}

\newcommand{\gtsimeq}{\raisebox{-0.6ex}{$\,\stackrel
         {\raisebox{-.2ex}{$\textstyle >$}}{\sim}\,$}}

\begin{document}

\title{The Pan-Pacific Planet Search. II. Confirmation of a two-planet 
system around HD\,121056}

\author{Robert A.~Wittenmyer\altaffilmark{1,2}, Liang 
Wang\altaffilmark{3}, Fan Liu\altaffilmark{4}, Jonathan 
Horner\altaffilmark{5,2}, Michael Endl\altaffilmark{6}, John Asher 
Johnson\altaffilmark{7}, C.G.~Tinney\altaffilmark{1,2}, 
B.D.~Carter\altaffilmark{5} }

\altaffiltext{1}{School of Physics, University of New South Wales, 
Sydney 2052, Australia}
\altaffiltext{2}{Australian Centre for Astrobiology, University of New 
South Wales, Sydney 2052, Australia}
\altaffiltext{3}{Key Laboratory of Optical Astronomy, National 
Astronomical Observatories, Chinese Academy of Sciences, A20 Datun Road, 
Chaoyang District, Beijing 100012, China}
\altaffiltext{4}{Research School of Astronomy \& Astrophysics, 
Australian National University, Cotter Road, Weston Creek, ACT 2611 
Australia}
\altaffiltext{5}{Computational Engineering and Science Research Centre, 
University of Southern Queensland, Toowoomba, Queensland 4350, 
Australia}
\altaffiltext{6}{McDonald Observatory, University of Texas at Austin, 1 
University Station C1400, Austin, TX 78712, USA }
\altaffiltext{7}{Harvard-Smithsonian Center for Astrophysics, Cambridge, 
MA 02138 USA}
\email{
rob@unsw.edu.au}

\shorttitle{HD 121056 two-planet system }
\shortauthors{Wittenmyer et al.}

\begin{abstract}

\noindent Precise radial velocities from the Anglo-Australian Telescope 
confirm the presence of a rare short-period planet around the K0 giant 
HD\,121056.  An independent two-planet solution using the AAT data shows 
that the inner planet has $P=89.1\pm$0.1 days, and 
m~sin~$i=1.35\pm$0.17\,\Mjup.  These data also confirm the planetary 
nature of the outer companion, with m~sin~$i=3.9\pm$0.6\,\Mjup\ and 
$a=2.96\pm$0.16\,AU.  HD\,121056 is the most-evolved star to host a 
confirmed multiple-planet system, and is a valuable example of a giant 
star hosting both a short-period and a long-period planet.

\end{abstract}


\section{Introduction}

With the discovery of more than 700 extrasolar planets by the 
radial-velocity method, and several thousand planet candidates by the 
\textit{Kepler} spacecraft \citep{borucki10, batalha13}, the past 20 
years have seen tremendous growth in our understanding of the startling 
diversity of planetary systems in the Solar neighborhood.  At the same 
time, planet-search efforts are now expanding into new realms of 
parameter space, seeking to understand how the detailed properties of 
planetary systems depend on the properties of their host stars.  At the 
low-mass end, M dwarfs are being targeted by a number of optical and 
near-infrared radial-velocity surveys searching for rocky and 
potentially habitable planets \citep[e.g.][]{endl06, q10, bean10, 
john10, suvrath12, barnes12, bonfils13, wright14}.  Meanwhile, 
virtually everything we know about planetary systems orbiting stars more 
massive than our Sun has come from taking advantage of stellar 
evolution.  High-mass stars on the main sequence are too hot and rotate 
too rapidly \citep{galland05} for the radial-velocity technique to work.  
By observing higher-mass stars when they evolve off the main sequence 
into subgiants and giants, precision velocity measurements become 
possible.  The stars expand and cool, presenting an abundance of narrow 
spectral absorption lines for accurate velocity determination.  This 
approach has been successfully used by several teams in recent years 
\citep[e.g.][]{setiawan03, hatzes05, sato05, johnson06b, doellinger07, 
n09, 47205paper}.

One result arising from these studies of evolved stars is a relative 
deficit of short-period planets, despite obvious selection biases in 
favor of detecting them.  Several thousand subgiant and giant stars are 
being monitored by both \textit{Kepler} and the programs mentioned 
above, but only seven planets are known to orbit giant stars 
(log\,$g<$3.5) within $a<0.5$\,AU.  They are HD\,102956b 
($a=0.0810$\,AU, Johnson et al.~2010), Kepler-91b ($a=0.072$\,AU, 
Lillo-Box et al.~2014), KOI-1299b ($a=0.3034$\,AU, Ciceri et al.~2014) 
and the two-planet systems orbiting Kepler-391 ($a_b=0.0744$\,AU, 
$a_c=0.1465$\,AU, Rowe et al.~2014) and Kepler-56 ($a_b=0.1028$\,AU, 
$a_c=0.1652$\,AU, Huber et al.~2013).  The apparent shortfall of 
planets orbiting evolved stars has been noted by \citet{johnson07} and 
\citet{sato10}.  Two possible explanations are that either the planets 
are absent, or are swallowed by the host star as it expands 
\citep{kunitomo11, villaver14}.

From 2009 to 2014, the Pan-Pacific Planet Search (PPPS) observed 170 
Southern Hemisphere subgiants and first-ascent giants 
\citep{47205paper}.  The PPPS targets are redder than those observed by 
most surveys \citep{mortier13} -- we have chosen stars with 
$1.0\le(B-V)\le\,1.2$, whereas other surveys enforce $(B-V)\le\,1.0$.  
This color selection makes the PPPS targets complementary to the 
$\sim$450 Northern ``retired A stars'' from the well-established Lick 
and Keck program \citep{johnson06, johnson11}.  A complete target list 
is given in \citet{47205paper}.  HD\,121056 (HIP\,67851, HR\,5224) is a 
bright Southern ($V=6.19$, RA 13:53:52.06, Dec -35:18:51.7) giant 
targeted by both the PPPS and the EXPRESS project \citep{jones11}, a 
similar radial-velocity campaign aimed at detecting planets orbiting 
Southern hemisphere evolved stars.  Recently, \citet{jones14} announced 
the discovery of two companions orbiting HD\,121056: a rare short-period 
planet at $P=88.8$\,d and an unconstrained massive outer body with an 
orbital period exceeding their observational baseline of 1523 days (4.2 
yr).

In this paper, we present independent data which confirm both the inner 
planet and the planetary nature of the outer body in the HD\,121056 
system.  Section~2 briefly describes the observational data and gives 
the stellar parameters.  In Section~3, we detail the orbit-fitting 
process and give the planetary parameters.  Finally, in Section~4 we 
give our conclusions and place this discovery in context.

\section{AAT Observations and Stellar Properties}

Precision Doppler measurements for the PPPS are obtained with the UCLES 
echelle spectrograph \citep{diego:90} at the 3.9-metre Anglo-Australian 
Telescope (AAT).  The observing procedure is identical to that used by 
the long-running Anglo-Australian Planet Search 
\citep[e.g.][]{tinney01,butler01,jones10,142paper}; a 
1-arcsecond slit delivers a resolving power of $R\sim$45,000.  
Calibration of the spectrograph point-spread function is achieved using 
an iodine absorption cell temperature-controlled at 
60.0$\pm$0.1$^{\rm{o}}$C.  The iodine cell superimposes a forest of 
narrow absorption lines from 5000 to 6200\,\AA, allowing simultaneous 
calibration of instrumental drifts as well as a precise wavelength 
reference \citep{val:95,BuMaWi96}.  Velocities are obtained using the 
\textit{Austral} code \citep{endl00}, which has been successfully used 
by several planet-search programs for more than 10 years 
\citep{endl04,texas1,sato13}.

We have obtained 22 observations of HD\,121056 since 2009 Feb 3, and an 
iodine-free template spectrum was obtained on 2010 July 4.  With 
$V=6.19$, exposure times are typically 300-600\,s, with a resulting S/N 
of $\sim$150-300 per pixel each epoch.  The data, given in 
Table~\ref{aatvels}, span a total of 1879 days (5.5\,yr), and have a 
mean internal velocity uncertainty of 4.0\,\ms.

We have used our iodine-free template spectrum ($R\sim$60,000, 
S/N$\sim$250) to derive spectroscopic stellar parameters.  In brief, the 
iron abundance [Fe/H] was determined from the equivalent widths of 32 
unblended Fe lines, and the LTE model atmospheres adopted in this work 
were interpolated from the ODFNEW grid of ATLAS9 \citep{Castelli2004}.  
The effective temperature ($T_\mathrm{eff}$) and bolometric correction 
($BC$) were derived from the color index $B-V$ and the estimated 
metallicity using the empirical calibration of 
\citet{Alonso1999,Alonso2001}.  Since the color-$T_\mathrm{eff}$ method 
is not extinction-free, we corrected for reddening and 
extinction using $E(B-V)=0.014$ and $A_V=3.1\times\,E(B-V)=0.043$ 
\citep{schlegel98}.  The stellar mass and age were estimated from the 
interpolation of Yonsei-Yale ($\mathrm{Y}^2$) stellar evolution tracks 
\citep{Yi2003}.  The resulting stellar mass of 1.30$\pm$0.18\Msun\ was 
adopted for calculating the planet masses.  Results from this method are 
labeled ``Method 1'' in Table~\ref{stellarparams}.


We also derived stellar parameters with the MOOG program 
\citep{sneden73}, based on the homogeneous, plane-parallel and local 
thermodynamic equilibrium models (1D-LTE) from \citet{cast03}.  The 
program matches the observed equivalent widths (EW) with theoretical 
values calculated based on the atmospheric model.  We obtained the 
effective temperature by forcing a consistent iron abundance, derived 
from 81 Fe I lines with their excitation potentials.  We determine 
log\,$g$ by forcing the Fe I and Fe II lines to give the same iron 
abundance.  The microturbulent velocity $v_t$ is determined by requiring 
a zero-slope relation between the log of the iron abundance and the EWs.  
The results from this method are labeled ``Method 2'' in 
Table~\ref{stellarparams}.

\section{Orbit Fitting and Planetary Parameters}

We fit for the planets in two ways: first, using only the AAT 
observations as a wholly independent check of \citet{jones14}, and 
then we include the published velocities for a joint solution.

\subsection{AAT/UCLES data only}

The 22 AAT velocities for HD\,121056 have an rms scatter of 64\ms\, and 
showed an obvious trend after $\sim$2 years.  The trend turned over in 
early 2013, and could be tentatively fit with a long-period object, 
though the rms scatter remained stubbornly high ($\sim$30\,\ms) compared 
to the mean velocity uncertainty of 4.0\,\ms\ for this bright star.  Due 
to the limited amount of data available, we used a genetic algorithm 
\citep{charbonneau95} rather than a traditional Lomb-Scargle periodogram 
search \citep{lomb76,scargle82}.  We have successfully used this 
technique in previous work \citep[e.g.][]{tinney11, 47205paper, NNSer} 
to detect planetary signals when data are sparse or when the candidate 
orbital periods are highly uncertain.  To check the results of 
\citet{jones14}, we allowed both planets to take on a very wide range of 
orbital periods ($P_{1}: 40-500$d, $P_{2}: 1200-2500$d), and we ran the 
genetic algorithm for 50,000 iterations, testing a total of about $10^7$ 
possible configurations.  Convergence rapidly occurred, on a solution in 
which the inner planet is consistent with the result of \citet{jones14}, 
and the outer planet has a period $P\sim$1650 days with low 
eccentricity.  

The best two-planet solution was then used as a starting point for the 
generalized least-squares program \textit{GaussFit} \citep{jefferys88}, 
here used to solve a Keplerian radial-velocity orbit model.  For the 
final fitting, we excluded the velocity point from JD 2456051, as it lay 
$5\sigma$ from the model and was obtained in exceedingly poor seeing 
($\sim$\,4\arcsec).  The rms about the 2-planet model was 6.4\ms, but 
dropped to 3.2\ms\ after removing that point.  The data and model fit 
are shown in Figure~\ref{fitplots}, and the best-fit system parameters 
are given in Table~\ref{planetparams}.  Since the formal 
uncertainties derived from the covariance matrix may be underestimated, 
we we also estimated parameter uncertainties using a bootstrap 
randomisation method within the \textit{Systemic Console} \citep{mes09}, 
which generates 10,000 simulated datasets by drawing with replacement 
from the original data.  Since the chief constraint on the outer planet 
comes from the first data point (at JD 2454866), the bootstrap method 
provides more realistic (and much larger) uncertainty estimates, 
particularly for the orbital period and mass.  The rightmost two columns 
of Table~\ref{planetparams} give the mean value and $1\sigma$ 
uncertainties for each parameter from the bootstrapping.

\subsection{Combined fit with published data}

We repeated the above fitting procedures, now including the 60 
velocities from FEROS and CHIRON as published in \citet{jones14}.  The 
results are given in Table~\ref{combinedfit}.  The rms scatter about the 
fit for the three data sets is as follows: AAT -- 9.26 \ms, FEROS -- 
11.02 \ms, CHIRON -- 9.85 \ms.  This scatter is considerably higher, but 
consistent with the expected pulsation-induced velocity jitter, as 
demonstrated by \citet{endl09} for Gamma Cephei, a giant star of similar 
evolutionary status as HD\,121056.  The combined fit gives a near-zero 
eccentricity for the inner planet and a somewhat longer period for the 
outer planet.  The bootstrap process favours a period near 2200 days 
(with a large uncertainty of 486 d), similiar to the 2100\,d estimate of 
\citet{jones14} and statistically consistent with the shorter periods 
given in Table~\ref{planetparams}.  Most importantly, all the fitting 
permuations described in this section unambiguously confirm the presence 
of the 89-day inner planet and the planetary nature of the outer 
companion, even when accounting for the high leverage of the first 
observation.

\section{Discussion}

\subsection{Evidence for orbiting planets}

\begin{figure}
\plottwo{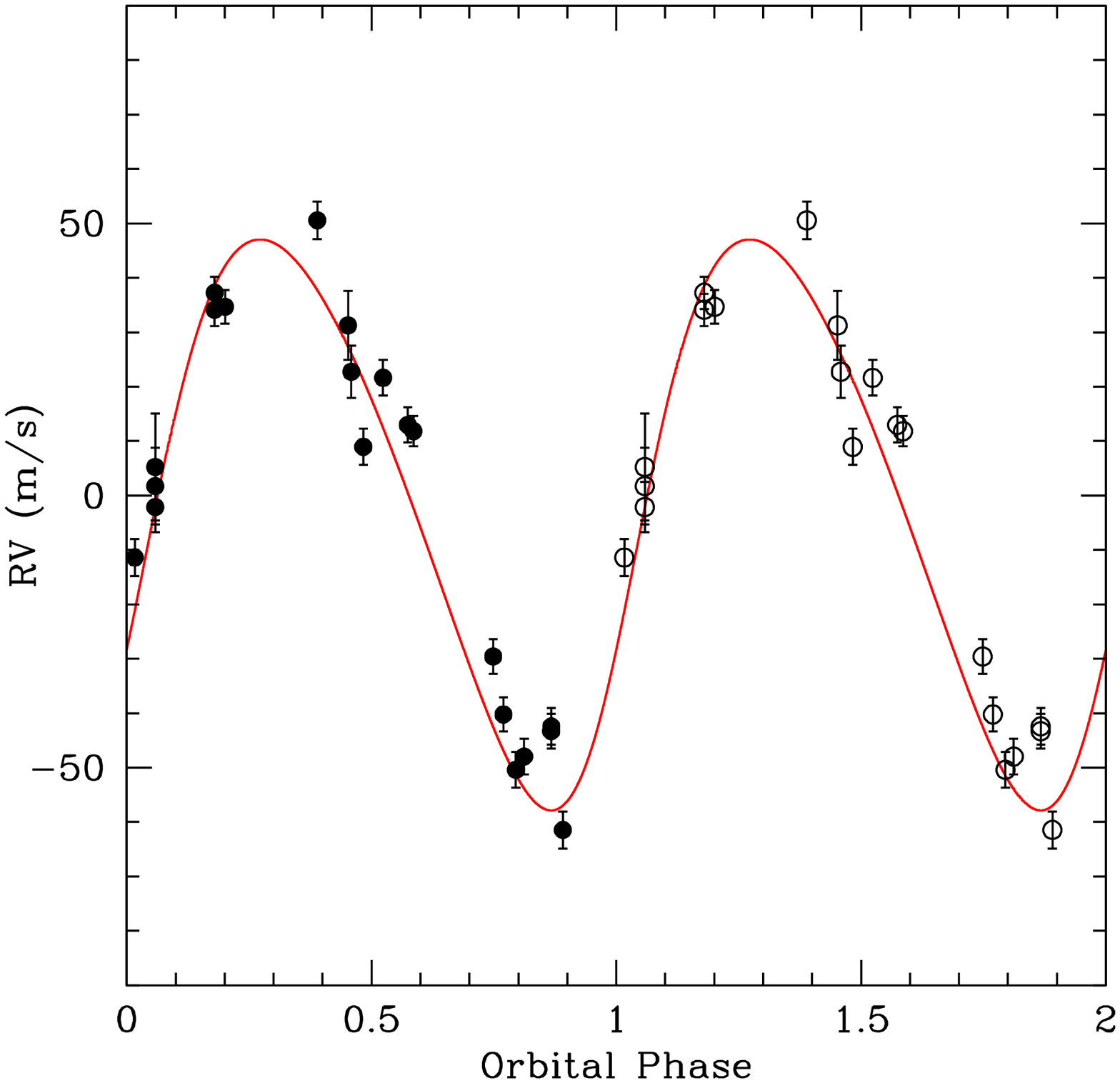}{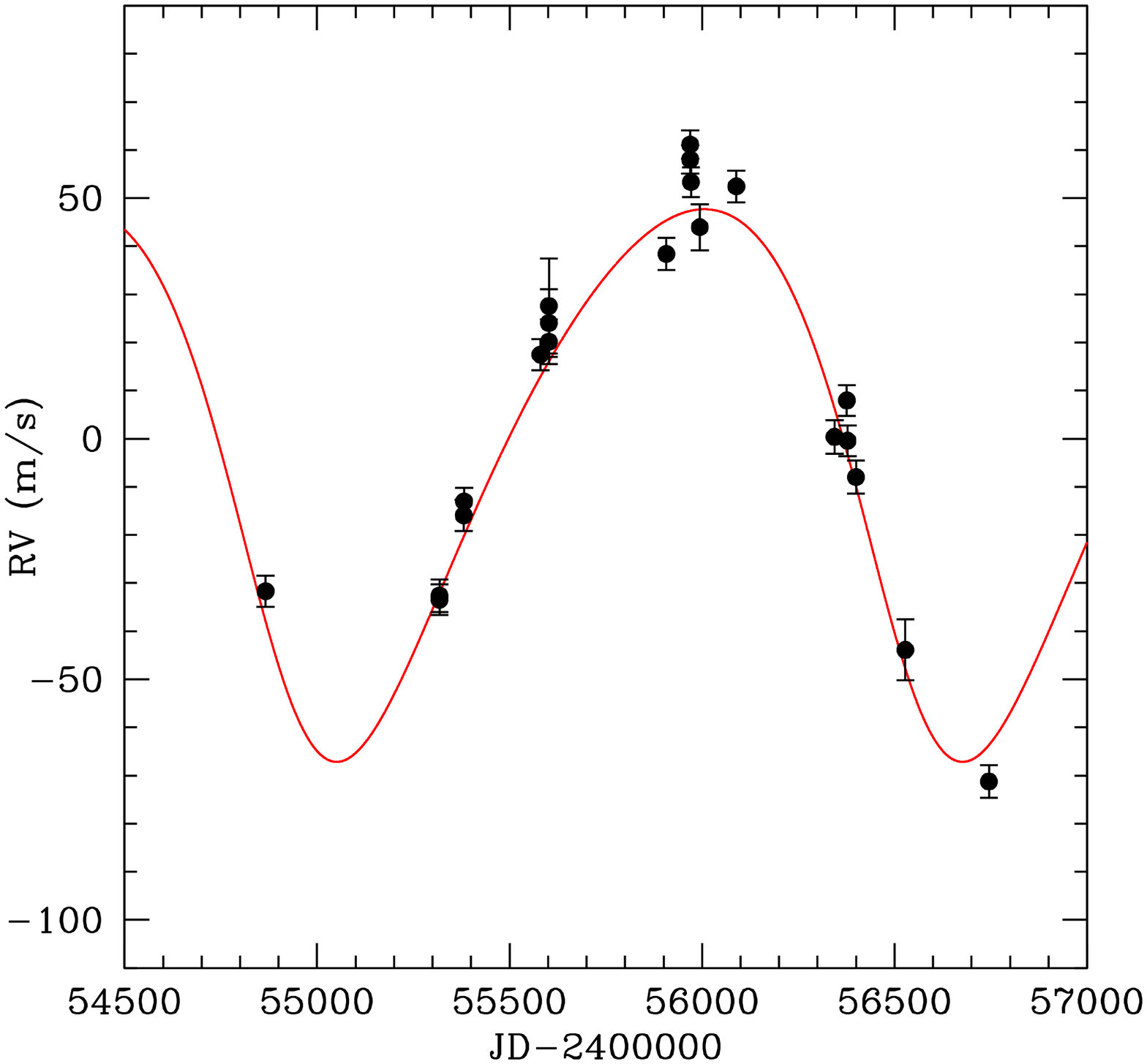}
\caption{AAT data and model fit for the HD\,121056 planets.  Left panel: 
Phase plot and model fit for HD\,121056b, with the outer planet removed. 
Two cycles are shown for clarity.  Right panel: Radial-velocity time 
series for HD\,121056c, with the inner planet removed.  The rms about 
the two-planet fit is 3.2\,\ms. }
\label{fitplots} 
\end{figure}

For any planet discovery, it is important to check for the possibility 
that the observed radial-velocity variations are intrinsic to the star 
(or the instrument) and not due to orbiting planets.  For HD\,121056, 
the periods of the two signals (89 and $\sim$1700 days) are nowhere near 
the window function peak at $\sim$21 days, nor are they near one year.  
Spurious periods in observational data most commonly arise at those 
periods due to sampling (imposed by the lunar cycle and yearly 
observability of a given target).  We also note that there are no large 
phase gaps in the data (Figure~\ref{fitplots}) -- such gaps would also 
cast doubt on the reality of a signal.  There are also no potentially 
contaminating background objects within 5 arcminutes of HD\,121056.

The chromospheric activity index $S_{HK}$, determined from observations 
of the Ca\,\textsc{II} H and K lines, is a critically important 
measurement for radial-velocity planet detection, as the star's activity 
is usually correlated with velocity variations that can mimic the reflex 
velocities induced by orbiting planets 
\citep[e.g.][]{bonfils07,hatzes10}.  \citet{jones14} demonstrated that 
the $S_{HK}$ index was not correlated to their velocities, supporting 
the planet hypothesis.  For radial-velocity detected planets, a major 
concern is that the signal is due to rotational modulation of starspots.  
For some stars, particularly giants, spots can induce quasi-periodic 
velocity variations in excess of 20\,\ms\ \citep{hekker08}.  For a 
spotted star, the rotation period can be deduced from photometry; 
\citet{jones14} found no significant periodicities in \textit{Hipparcos} 
photometry, and no correlations between the radial velocities and the 
line profiles as measured in the bisector velocity span and the 
cross-correlation FWHM.  Following \citet{47205paper}, we can combine 
the available estimates of the star's radius and v~sin~$i$ 
(Table~\ref{stellarparams}) minimum rotational velocity to obtain a 
maximum rotation period.  This calculation yields a \textit{maximum} 
rotation period of 167$\pm$83 days, which is distinctly different 
from either candidate planet's orbital period.  Furthermore, if the 
89-day signal were caused by spots, it seems exceedingly unlikely that 
one or more starspots would persist so coherently over the 5.5 years of 
our observations such that we detect the clean and well-sampled signal 
presented here (Figure~\ref{fitplots}).  These lines of evidence lead us 
to conclude that the simplest explanation for the observed highly 
significant and coherent radial-velocity variations is the gravitational 
influence of two orbiting giant planets.

\subsection{Dynamical considerations}

There is a growing body of recent work calling for the rigorous 
dynamical stability testing of proposed multiple-planet systems 
\citep[e.g.][]{horner11, hinse12, NSVS, QSVir}.  Detailed dynamical 
testing can confirm or refute the orbital configurations inferred from 
Keplerian fits, and hence is a valuable part of the discovery process.

In many cases, when the planets proposed are relatively tightly packed, 
dynamically speaking, the only way to address the question of their 
stability is to run large scale suites of dynamical simulations, such as 
those presented in \citet{songhu}.  In some cases, however, the proposed 
planets are so widely spaced that it is highly unlikely that they would 
interact with one another sufficiently strongly to disrupt the system. 
Such planets are essentially decoupled from one another 
(e.g.~HD\,159868b,c: Wittenmyer et al.~2012).

\citet{gladman93} found that, for planets moving on orbits with low 
eccentricity and inclination, two planet systems are typically stable 
when the orbits of the planets are separated by more than $2\sqrt{3}$ 
times their mutual Hill radius, where the mutual Hill radius is defined 
as:

\begin{equation}
R_H = \Big{[}\frac{(m_1 + m_2)}{3 \Msun}\Big{]}^{1/3} \Big{[}\frac{(a_1 + a_2)}{2}\Big{]} .
\end{equation}

\noindent \citet{chambers96} tested this using numerical integrations, 
and confirmed that this result holds, in general.  As such, this seems a 
reasonable first criterion by which the potential stability of which a 
given two-planet system can be considered.  Clearly, the greater the 
separation of a given two planet system, measured in Hill radii, the 
more likely it is to be truly dynamically stable.

In the case of the HD\,121056 system, Equation (1) reveals that the 
mutual Hill radius of the two planets proposed in this work is 
$R_{H}\sim$\,0.265 AU.  The orbits of the planets are thus separated by 
$\sim$9.6 mutual Hill radii - far more widely spaced than the 
$2\sqrt{3}R_{H}$ stability criterion of \citet{gladman93}.  Indeed, the 
planets are sufficiently widely spaced that the system is likely to have 
dynamical room for at least one additional planet between their orbits.  
Our AAT data, with a residual rms scatter of only 3.2\ms, can be used to 
place reasonably tight limits on the presence of any undetected planet.  
In brief, we add the Keplerian velocity signal of a fictitious 
planet to the residuals of our fit, and attempt to recover it via a 
generalized Lomb-Scargle periodogram \citep{zk09}.  Here, we have 
assumed circular orbits; for each combination of period $P$ and 
radial-velocity semiamplitude $K$, we tried 30 values of orbital phase.  
A planet is deemed detectable if 99\% of orbital configurations at a 
given $P$ and $K$ are recovered with a false-alarm probability 
\citep{ss10} of less than 1\%.  This approach is essentially identical 
to that used in our previous work \citep[e.g.][]{limitspaper, jupiters, 
pasp}.  Using the AAT-only fit residuals, we can with 99\% confidence 
exclude planets with m~sin~$i>0.41$\,\Mjup\ on circular orbits between 
the orbital excursions of the two planets ($0.5<a<2.3$\,AU).  Including 
all available data as in \S 3.2, we obtain a mass limit of 
m~sin~$i>0.66$\,\Mjup.  Further observations over additional orbital 
cycles of the outer planet will substantially improve the quality of the 
2-Keplerian fit and better constrain the region between the planets.

\subsection{Conclusions}

\begin{figure}
\plotone{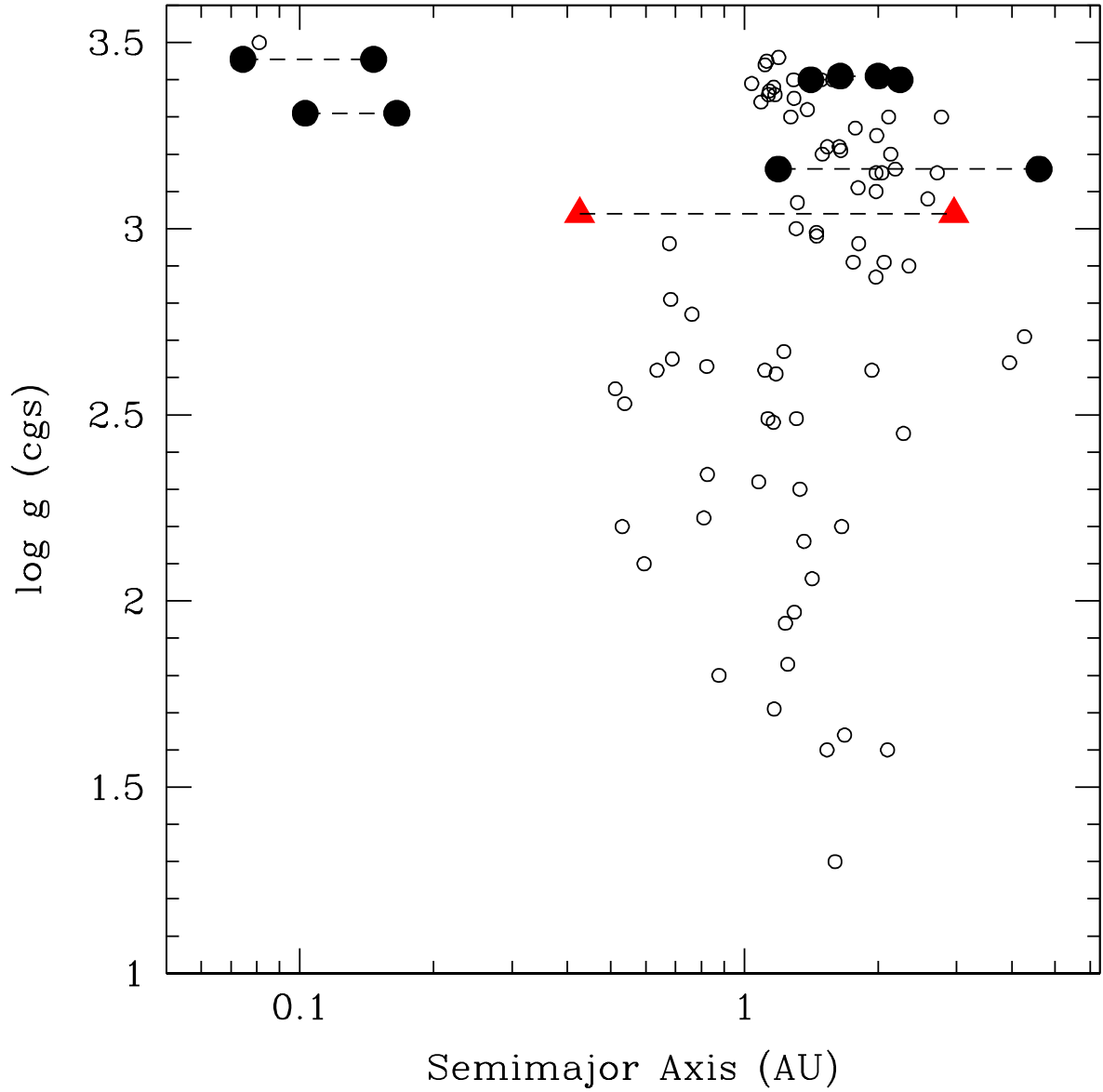}
\caption{Confirmed planets orbiting giant stars (log $g<$3.5).  Planets 
in multiple systems are shown as filled circles connected by a dashed 
line.  BD+20~2457 \citep{n09b} is not shown as it was demonstrated to be 
dynamically unfeasible by \citet{bdpaper}.  HD\,121056b and c are shown 
as red triangles. }
\label{theymightbegiants}
\end{figure}

We have independently confirmed the existence of a system of two giant 
planets orbiting the K0 giant HD\,121056 \citep{jones14}.  Such 
confirmation is essential as the system hosts a close-in planet 
($a<0.5$\,AU), a rarity among giant stars.  The longer baseline of our 
data also confirm the planetary nature of the outer companion proposed 
by \citet{jones14}.  The multiplicity of the HD\,121056 system makes it 
stand out from the growing crowd of planets known to orbit evolved 
stars.  Only six multiple-planet systems have been found around giant 
stars: HD\,4732 \citep{sato13}, HD\,200964 and 24 Sex 
\citep{johnson11a}, Kepler-391 \citep{rowe14}, Kepler-56 
\citep{huber13}, and BD+20~2457 \citep{n09b}.  However, the latter 
system has been shown to be dynamically unstable \citep{bdpaper}, 
casting doubt on its veracity.  If we consider the surface gravity 
log~$g$ as a proxy for the degree to which a star has evolved off the 
main sequence, HD\,121056 is the most-evolved star to host a 
multiple-planet system (Figure~\ref{theymightbegiants}).  With 
this in mind, the lower eccentricity for HD\,121056b found by the 
combined fit (Table~\ref{combinedfit}) seems more physically plausible, 
arising from tidal circularization interactions with the expanding host 
star.  The fate of planets as their host stars evolve have been modeled 
in detail \citep[][e.g.]{vl09,kunitomo11,mustill14}.  Unsurprisingly, at 
$a=0.426$\,AU, HD\,121056b is doomed to be engulfed.  The fate of 
HD\,121056c is less clear: models by \citet{vl07} show that for a 
1\Msun\ star, planets beyond $a\sim$3 AU are likely to survive the 
asymptotic giant branch phase. However, \citet{vl09} showed that a 
5\Mjup\ planet must have $a\gtsimeq$3.7\,AU to avoid tidal capture in 
the red-giant phase.  \citet{mustill12} likewise find the minimum 
orbital distance for survival to be $\sim$2.6\,AU, increasing for 
eccentric planets.  The significant uncertainties in the eccentricity 
and semimajor axis of HD\,121056c mean that its fate rests on a knife 
edge, to be crystallised by future observations refining its orbit. 

The PPPS aimed to explore the dependence of planetary system 
properties on host-star mass by improve the detection statistics for 
intermediate-mass stars (1.5-3.0 \Msun).  Many of the other programs 
listed in \S 1 are likewise targeting evolved stars to explore the same 
parameter space -- though the masses of such stars have been put into 
question \citep[][e.g.]{lloyd11, lloyd13, johnson13, johnson14}.  While 
HD\,121056 is most likely a near-solar-mass star 
(Table~\ref{stellarparams}), its two super-Jovian mass planets are 
characteristic of those planets orbiting higher-mass stars.  That is, 
planets orbiting stars more massive than the Sun tend to be more massive 
\citep{bowler10} and move on more circular orbits \citep{johnson08}.

By good fortune, our data for HD\,121056 sampled the inner planet's 
phase well enough to enable secure detection with relatively few 
observations.  \citet{jones14} were able to obtain high-cadence CHIRON 
observations to confirm the inner planet.  This example highlights the 
need for high-cadence monitoring of evolved stars to determine whether 
the deficit of close-in planets (e.g. Figure~\ref{theymightbegiants}) is 
real or an observational bias arising from highly competitive and 
sparsely-scheduled time on large telescopes.  Dedicated exoplanet 
observatories such as the Automated Planet Finder \citep{vogt14} or 
MINERVA \citep{minerva, swift14} are the way forward.  Further 
opportunities can be found with smaller telescopes equipped with 
high-resolution spectrographs for precise Doppler velocimetry, e.g. New 
Zealand's Mount John University Observatory \citep{hearnshaw02,endl14} 
and Shandong University's Weihai Observatory \citep{gao14, cao14}.

\acknowledgements

CGT is supported by Australian Research Council grants DP0774000 and 
DP130102695.  JH and BDC are supported by USQ's Strategic Research Fund.  
We gratefully acknowledge the efforts of PPPS guest observers Hugh Jones 
and Simon O'Toole.  This research has made use of NASA's Astrophysics 
Data System (ADS), and the SIMBAD database, operated at CDS, Strasbourg, 
France.  This research has also made use of the Exoplanet Orbit Database 
and the Exoplanet Data Explorer at exoplanets.org \citep{wright11}.


\begin{deluxetable}{lrr}
\tabletypesize{\scriptsize}
\tablecolumns{3}
\tablewidth{0pt}
\tablecaption{AAT Radial Velocities for HD 121056}
\label{aatvels}
\tablehead{
\colhead{JD-2400000} & \colhead{Velocity (\ms)} & \colhead{Uncertainty
(\ms)}}
\startdata
54866.26766  &    -86.6  &    3.3  \\
55318.00434  &    -87.0  &    3.2  \\
55318.00981  &    -86.2  &    3.4  \\
55380.96409  &    -19.8  &    3.2  \\
55382.00821  &    -20.7  &    2.8  \\
55580.24855  &    -38.0  &    3.3  \\
55602.20623  &     19.8  &    9.8  \\
55602.21786  &     16.3  &    7.1  \\
55602.22211  &     12.5  &    4.6  \\
55907.24206  &     63.0  &    3.3  \\
55969.24861  &     87.6  &    3.0  \\
55969.26001  &     90.7  &    2.9  \\
55971.15907  &     88.1  &    3.1  \\
55994.12084  &     75.1  &    4.8  \\
56051.94440\footnote{Not included in fit.}  &     32.8  &    4.0  \\
56088.95971  &     65.0  &    3.3  \\
56344.18468  &     45.0  &    3.5  \\
56376.17506  &    -42.2  &    3.2  \\
56378.04416  &    -53.2  &    3.1  \\
56400.03556  &    -29.2  &    3.4  \\
56527.88115  &    -11.1  &    6.3  \\
56745.10350  &   -122.1  &    3.4  \\
\enddata
\end{deluxetable}

\begin{deluxetable}{lllll}
\tabletypesize{\scriptsize}
\tablecolumns{5}
\tablewidth{0pt}
\tablecaption{Stellar Parameters for HD 121056}
\tablehead{
\colhead{Parameter} & \colhead{Method 1} & \colhead{Method 2} & 
\colhead{Literature} & \colhead{Reference}
 }
\startdata
\label{stellarparams}
Spec.~Type & & & K0 III & \citet{houk82} \\
Distance (pc) & & & 66.0$\pm$1.7 & \citet{vanl07} \\
$(B-V)$ & & & 1.008$\pm$0.014 & \citet{vanl07} \\
Mass (\Msun) & 1.30$\pm$0.18 & 1.21 & 1.63$\pm$0.22 & \citet{jones14} \\
  & & &  1.4 & \citet{randich99} \\
V sin $i$ (\kms) & $<$2 & & 1.4 & \citet{randich99} \\
  & & & 1.8$\pm$0.9 & \citet{jones11} \\
$[Fe/H]$ & -0.03$\pm$0.10 &  -0.13$\pm$0.06 & -0.11$\pm$0.09 & \citet{randich99} \\
  & & &  0.00$\pm$0.10  & \citet{jones11} \\
$T_{eff}$ (K) & 4805$\pm$100 & 4859$\pm$100 & 4711$\pm$100 & \citet{randich99} \\
  & & & 4890$\pm$100 & \citet{jones14} \\
log $g$ & 3.04$\pm$0.10 & 2.89$\pm$0.10 & 3.0$\pm$0.3 & \citet{randich99} \\
  & & & 3.15$\pm$0.20 & \citet{jones14} \\
$v_t$ (\kms) & 1.35$\pm$0.09 & 1.21$\pm$0.08 & & \\
Radius (\Rsun) & 5.87$\pm$0.29 & & 5.92$\pm$0.44 & \citet{jones14} \\
Luminosity (\Lsun) & 16.1$\pm$0.8 & 18.2 & 17.6$\pm$2.6 & \citet{jones14} \\
\enddata
\end{deluxetable}

\begin{deluxetable}{lllll}
\tabletypesize{\scriptsize}
\tablecolumns{5}
\tablewidth{0pt}
\tablecaption{HD\,121056 Planetary System Parameters (AAT data only) }
\tablehead{
\colhead{Parameter} & \multicolumn{2}{c}{\textit{GaussFit} Solution} & 
\multicolumn{2}{c}{Bootstrap Solution\tablenotemark{a}} \\
\colhead{} & \colhead{HD\,121056b} & \colhead{HD\,121056c} & 
\colhead{HD\,121056b} & \colhead{HD\,121056c}
}
\startdata
\label{planetparams}
Period (days) & 89.06$\pm$0.10 & 1626$\pm$26 & 88.99$\pm$0.33 & 1653$\pm$132 \\
Eccentricity & 0.17$\pm$0.04 & 0.20$\pm$0.05 & 0.34$\pm$0.19 & 0.34$\pm$0.18 \\
$\omega$ (degrees) & 244$\pm$10 & 136$\pm$11 & 213$\pm$38 & 155$\pm$54 \\
$K$ (\ms) & 52.5$\pm$2.2 & 57.4$\pm$3.1 & 60.2$\pm$11.0 & 70.5$\pm$17.5 \\
$T_0$ (JD-2400000) & 55329.8$\pm$2.6 & 53290$\pm$59 & 54785$\pm$12 & 53308$\pm$248 \\
m sin $i$ (\Mjup) & 1.35$\pm$0.17 & 3.88$\pm$0.55 & 1.48$\pm$0.30 & 4.60$\pm$1.37 \\
$a$ (AU) & 0.426$\pm$0.020 & 2.96$\pm$0.16 & 0.426$\pm$0.021 & 2.99$\pm$0.27 \\
\hline
RMS of fit (\ms) & 3.23 &   & 3.16 &   \\
$\chi^2_{\nu}$ & 1.36 &  & 1.36 &  \\
\enddata
\tablenotetext{a}{Mean parameter value and 68.7\% confidence interval 
from 10,000 bootstrap iterations.}
\end{deluxetable}


\begin{deluxetable}{lllll}
\tabletypesize{\scriptsize}
\tablecolumns{5}
\tablewidth{0pt}
\tablecaption{HD\,121056 Planetary System Parameters (all data) }
\tablehead{
\colhead{Parameter} & \multicolumn{2}{c}{\textit{GaussFit} Solution} & 
\multicolumn{2}{c}{Bootstrap Solution\tablenotemark{a}} \\
\colhead{} & \colhead{HD\,121056b} & \colhead{HD\,121056c} & 
\colhead{HD\,121056b} & \colhead{HD\,121056c} }

\startdata
\label{combinedfit}
Period (days) & 89.09$\pm$0.12 & 1741$\pm$39 & 89.09$\pm$0.11 & 2203$\pm$486 \\
Eccentricity & 0.02$\pm$0.04 & 0.20$\pm$0.04 & 0.06$\pm$0.04 & 0.18$\pm$0.07 \\
$\omega$ (degrees) & 211$\pm$103 & 201$\pm$9 & 300$\pm$132 & 205$\pm$17 \\
$K$ (\ms) & 47.9$\pm$1.8 & 62.8$\pm$2.8 & 47.9$\pm$11.0 & 81.9$\pm$14.0 \\
$T_0$ (JD-2400000) & 55500$\pm$26 & 53427$\pm$56 & 54810$\pm$33 & 53068$\pm$405 \\
m sin $i$ (\Mjup) & 1.25$\pm$0.16 & 4.34$\pm$0.59 & 1.25$\pm$0.04 & 6.14$\pm$1.99 \\
$a$ (AU) & 0.426$\pm$0.020 & 3.09$\pm$0.18 & 0.426$\pm$0.020 & 3.62$\pm$0.58 \\
\hline
RMS of fit -- AAT (\ms) & 9.26 &   & 9.27 &   \\
RMS of fit -- FEROS (\ms) & 11.02 &   & 11.01 &   \\
RMS of fit -- CHIRON (\ms) & 9.85 &   & 9.85 &   \\
$\chi^2_{\nu}$ & 5.46 &  & 5.46 &  \\
\enddata
\tablenotetext{a}{Mean parameter value and 68.7\% confidence interval
from 10,000 bootstrap iterations.}
\end{deluxetable}


\begin{figure}
\plotone{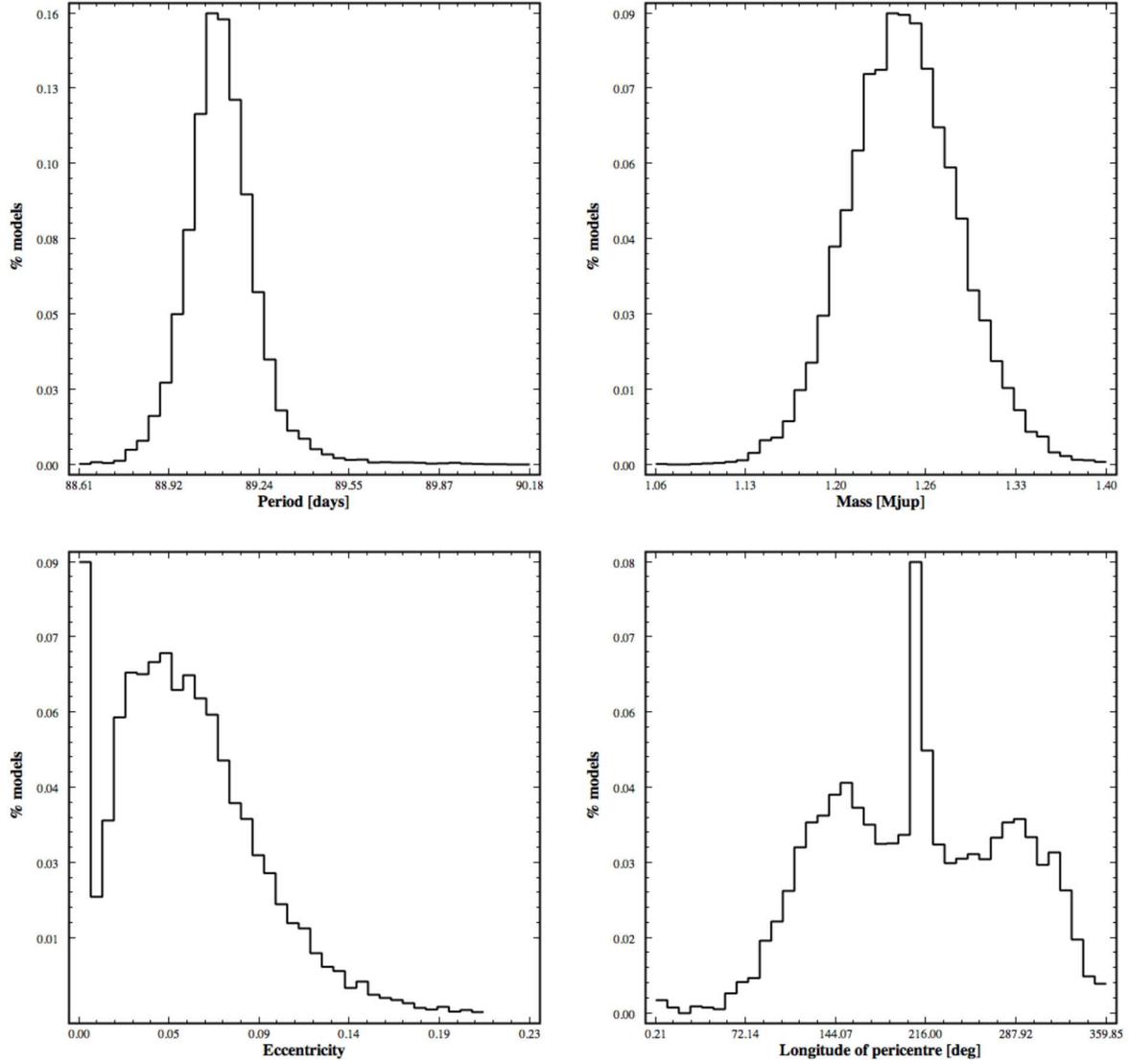}
\caption{Histogram of parameter distributions resulting from 10,000 
bootstrap iterations for HD\,121056b's orbital period, m~sin~$i$, 
eccentricity, and periastron argument $\omega$.  All velocity data were 
used. }
\label{pdf1} 
\end{figure}


\begin{figure}
\plotone{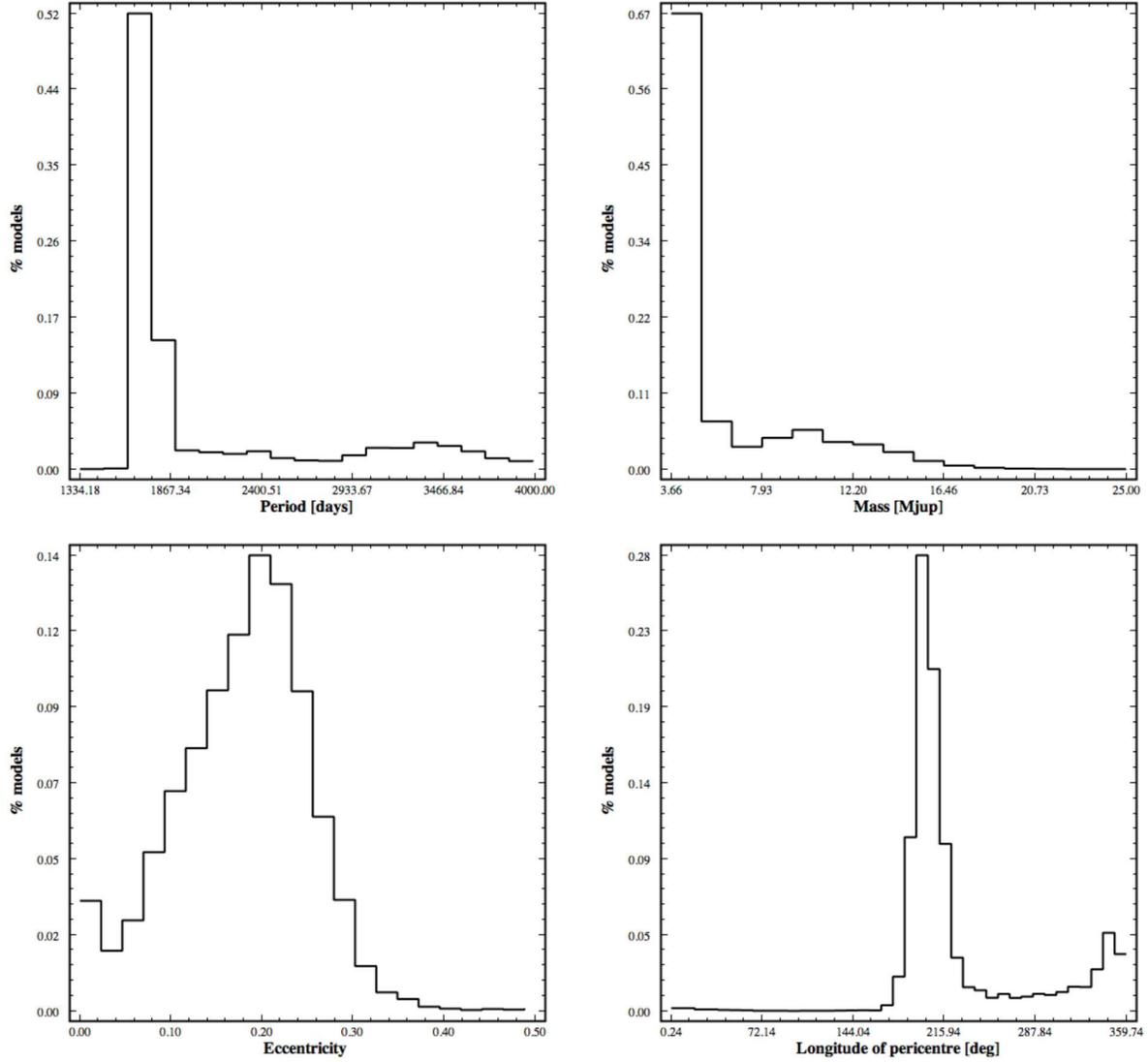}
\caption{Same as Figure~\ref{pdf1}, but for HD\,121056c. The best fit 
period for the outer planet relies critically on the first epoch, as 
demonstrated by the long tail in the period and mass distributions. }
\label{pdf2}
\end{figure}


\end{document}